\documentclass[preprint,linenumbers,showpacs,preprintnumbers,amsmath,amssymb,showkeys,aps,prd,a4paper,byrevtex]{revtex4}

\usepackage[dvips]{graphicx}
\usepackage{dcolumn}
\usepackage{bm}
\usepackage{epsfig}
\usepackage{amsmath}
\usepackage{amsfonts}
\usepackage{amssymb,amscd}
\usepackage[latin1]{inputenc}
\usepackage{hyperref}
\usepackage{braket}
\usepackage{url}
\usepackage{hepunits}
\usepackage{hepparticles}

\begin{document}

\title{Production of magnetic monopoles and monopolium in peripheral collisions}

\author{J.~T. Reis}
\email{jeantreis@gmail.com}
\author{W.~K. Sauter}
\email{werner.sauter@ufpel.edu.br}
\affiliation{Grupo de Altas e M\'edias Energias \\ Departamento de F\'{\i}sica - Instituto de F\'{\i}sica e Matem\'atica \\ Universidade Federal de Pelotas}

\date{\today}

\begin{abstract}
The exclusive production by the photon fusion of magnetic monopoles and the bound states of magnetic monopoles, the monopolium, is consider in different high energy processes. More specifically, we calculate the total cross sections of the ultraperipheral elastic collisions of electron-electron, proton-proton and lead-lead in present and future colliders, comparing with the previous results found in the literature. Our results indicates that magnetic monopoles or his bound states, if both exists, can be measuareble in future electron-electron colliders.
\end{abstract}

\maketitle

\section{Introduction \label{sc:in}}

With the discover of the Higgs boson\cite{Chatrchyan:2012xdj, Aad:2012tfa}, the last remaining unobserved particle of Standard Model (SM) was found. The attention now comes to the seek of signals of particles that not included in SM. Among of these (several) beyond Standard Model (BSM) new particles includes the radion~\cite{goncalves2010RadproexcproCERLHC}, a particle related tho Randall-Sundrum scenario of large extra dimensions and the dilaton, a BSM particle related which is a pseudo-Nambu-Goldstone boson in spontaneous breaking of scale symmetry\cite{Goncalves:2015oua} and also signals of extra dimensions~\cite{thiel2013HeaquaproblaholevaLHC,goncalves2014DifZbospaiproLHClarextdimsce}. 

An another predicted BSM particle, the magnetic monopole proposed by Dirac\cite{dirac1931QuaSinEleFie} gives a natural way to explain the quantization of electric charge. The magnetic monopole are also predicted in Grand Unified Theories (GUT)\cite{tHooft:1974kcl, Polyakov:1974ek}. Unfortunately, the predicted mass is very large and, until now the several  experimental searches not confirm his existence, only experimental limits on his mass and charge. Recently, the MOEDAL\cite{acharya2014PhyProMoEExpLHC}, a dedicated experiment on highly ionizating exotic particles is set on LHC and announces his first results\cite{MoEDAL:2016jlb,Acharya:2016ukt}. An revision of the state-of-art of theoretical and experimental status of magnetic monopole can be found in \cite{olive2014RevParPhy}.

In this work, we analyze the production of magnetic monopoles and the bound state, the monopolium\cite{Hill:1982iq}, by photon fusion in peripheral hadronic (proton-proton and ion-ion) collisons in present high energies at LHC and also in electron-positron expected at CLIC (see \cite{Aicheler:2012bya}). In particular, we consider the central exclusive production: the projectiles does not dissociate and the particle is produced in the central region of rapidity of the detector, giving a clean experimental signal of this process. 

We compare the our results with the previous ones in $pp$ collisions \cite{kurochkin2006promagmonviagamgamfushigeneppcol,dougall2009Dirmagmonprophofusprocol,epele2012LoomagmonLHCdipeve} and we present a prediction for the production of pairs of monopole/antimonopole and monopolium in $Pb-Pb$ collisions at LHC as well as in electron/positron collisions in the planned CLIC. 

This paper are organized as follows. In next session, we present a overview of the theory of magnetic monopoles with a short review of cross section production of a pair of monopoles and monopolium. In section \ref{sc:fpc}, we present the mechanism of central production in peripheral collisions with the central system of particles produced by a pair of high energy photons. The results of the calculation are showed and discussed in the section \ref{sc:res}. Finally, a summary and the conclusions are presented in the section \ref{sc:cls}. 

\section{Magnetic monopoles and monopolium \label{sc:mmr}}

Since the inception of the Classical Electrodynamics, is clear that the Maxwell Equations are not symmetric in relation of the electric charges. The possibility of existence of isolated magnetic charges, the magnetic monopoles, have interesting consequences, both in classical and quantum levels\cite{milton2006Theexpstamagmon,rajantie2012MagMonFieTheCos}. Probably, the most important is the Dirac quantization charge relation, which established that, if the magnetic charge exists, then the electric charge is quantized. The possibility of magnetic monopoles was found in Grand Unified Theories (GUT)\cite{tHooft:1974kcl, Polyakov:1974ek}. A review of the state-of-art of magnetic monopoles is found in \cite{olive2014RevParPhy}. See \cite{Patrizii:2015uea} for a recent review of experimental searches of this particles in colliders and cosmic rays. Meanwhile, the experimental difficulties for the experimental observation of isolated magnetic monopoles suggest the existence of a bound state, the monopolium\cite{
vento2008HidDirMon,epele2008MonKeytomon}.

We revisit the results of production of monopoles in $pp$ collisions at LHC energies\cite{ginzburg1998SeaheamagmonTevCERLHC,ginzburg1999Viseffverheamagmoncol,kurochkin2006promagmonviagamgamfushigeneppcol,dougall2009Dirmagmonprophofusprocol} and investigate the same process in ion-ion collisons in LHC and electron-electron collisions at CLIC. The production in nuclear collisions was proposed in \cite{roberts1986DirMagMonPaiProRelNucNucCol,he1997SeaDirmagmonhignuccol} in context of thermal quark-gluon plasma. In this context, in \cite{gould2017Magmonmasbouheaioncolandneusta} calculate bounds in the monopole mass from heavy ion collisions and neutron stars. Ginzburg and Schiller\cite{ginzburg1998SeaheamagmonTevCERLHC} previously consider the monopole pair production in electron-positron collisions for planned colliders and for photon luminosity of \cite{budnev1975TwophoparpromecPhyproAppEquphoapp}.

One of the ingredients of the calculation is the cross section of the process of fusion of two photons into a monopole/antimonopole pair or a monopolium. Unfortunately, due the high values of the coupling, a true perturbative calculation for all energies is questionable and therefore, the results presented here can be seen as an estimation for the cross sections. 

The coupling of the monopoles with photons can be quantified by two different forms. First, from Dirac himself, takes the coupling constant as $\alpha_{mag} = g^2/4\pi$ whereas the so-called beta coupling, consider $\alpha_{mag} = (\beta g)^2/4\pi$ where $\beta$ is the speed of monopole (in natural units)\cite{milton2006Theexpstamagmon}. We will use the Dirac expression for our results.

In the case of production of a antimonopole-monopole pair, the cross section can be obtained from the QED fundamental process of annihilation of a lepton pair into a pair of photons, changing the relevant physical quantities. In the center of mass frame, reads as
\begin{equation}
{\sigma}^{(\bar m m)}_{\gamma\gamma} =  \frac{\pi g^4(1-\beta^2)\beta^4}{2m^2}\left[ (3 - \beta^4)\ln\left(\frac{1 + \beta^2}{1 - \beta^2} \right) -2\beta(2-\beta^2) \right] . 
\label{eq:mn}
\end{equation}

For the monopolium production, we use the known result of cross section for the production of a massive ressonance,
\begin{equation}
\sigma(\gamma\gamma \rightarrow M) = \frac{4\pi}{\hat{s}}\frac{M^2\Gamma(\sqrt{\hat{s}})\Gamma_M}{(\hat{s} - M^2)^2 +M^2\Gamma^2_M},
\label{sec_monopolium1}
\end{equation}
where $M = 2m + E_{bound}$ is the monopolium mass, $\Gamma_M = \unit{10}{\GeV}$\cite{epele2008MonKeytomon} and
\[
 \Gamma(\sqrt{\hat{s}}) = \frac{8 \pi \alpha_{mag}^{2}}{m^2}|\psi_M(0)|^2.
\]
with $\psi_M(0)$ is the value of the wave function in the origin of the bound system of monopole/antimonopole. In this case, the pair are a bound state and are described as a massive resonance, characterized by your mass and decay widths. 

Related with the above process, an experimental signal of the monopolium production is the two photon production with the monopolium as a massive ressonance state, $\gamma\gamma \rightarrow M \rightarrow \gamma\gamma$. A possible background for this process is the production of two photons by a loop of leptons/quarks, which cross section can be estimated by results from the Standard Model. 

The values of $\psi(0)$ and $E_{bound}$ are obtained for the solution of the radial Schr\"odinger equation for the Coulomb like potential\cite{epele2008MonKeytomon}
\begin{equation}
 V(r)  \simeq -\frac{g^2}{4\pi}\frac{1}{r}.
\end{equation}
We use the solutions for this potential for the energy eigenvalues,
\begin{equation}
 E_n = -\left(\frac{1}{8\alpha_{elm}}\right)^2\frac{m}{n^2}
\end{equation}
 and the value of wave eigenfunctions in the origin,
\begin{equation}
 \psi_{n00} = \frac{1}{\sqrt{\pi}}\left(\frac{m}{8\alpha_{elm}n}\right)^{3/2}.
\end{equation}
The condition of bound state, $0\leq 2m + E_n \leq 2m$, imposes a condition on allowed values of $n$. With the above eigenvalues, only values $n\geq 13$ are possible and we will use $n=13$.

Recently, \cite{Barrie:2016wxf_2} proposed a monopolium model with finite size based on the 't Hooft-Polyakov solution and $U(1)$ lattice gauge theory which result on a binding potential with a linear term, similar to the Cornell potential of the quarkonium states in QCD\cite{Eichten:1978tg}. A similar approach was also proposed in \cite{saurabh2017MonIntPot}. In a future work, we will consider the numerical solutions for the eigenstates of monopolium for this class of confining potentials.

\section{Formalism of peripheral collisions \label{sc:fpc}}

The process of central production of particles\cite{khoze2001DoudifprohigresmismasexpTev,khoze2000CanHigbeseerapgapeveTevorLHC} has attracted much attention in the recent years, specially with the start of the LHC operation and besides the dedicated experiments to his observation\cite{albrow2009FP4ampProHigNewPhyforproLHC}. Beyond the Higgs boson, several other particles can be produced, some in a expressive ratio, inside this mechanism.

\begin{figure}[ht] 
 \begin{center}
 \includegraphics*[width=0.45\columnwidth]{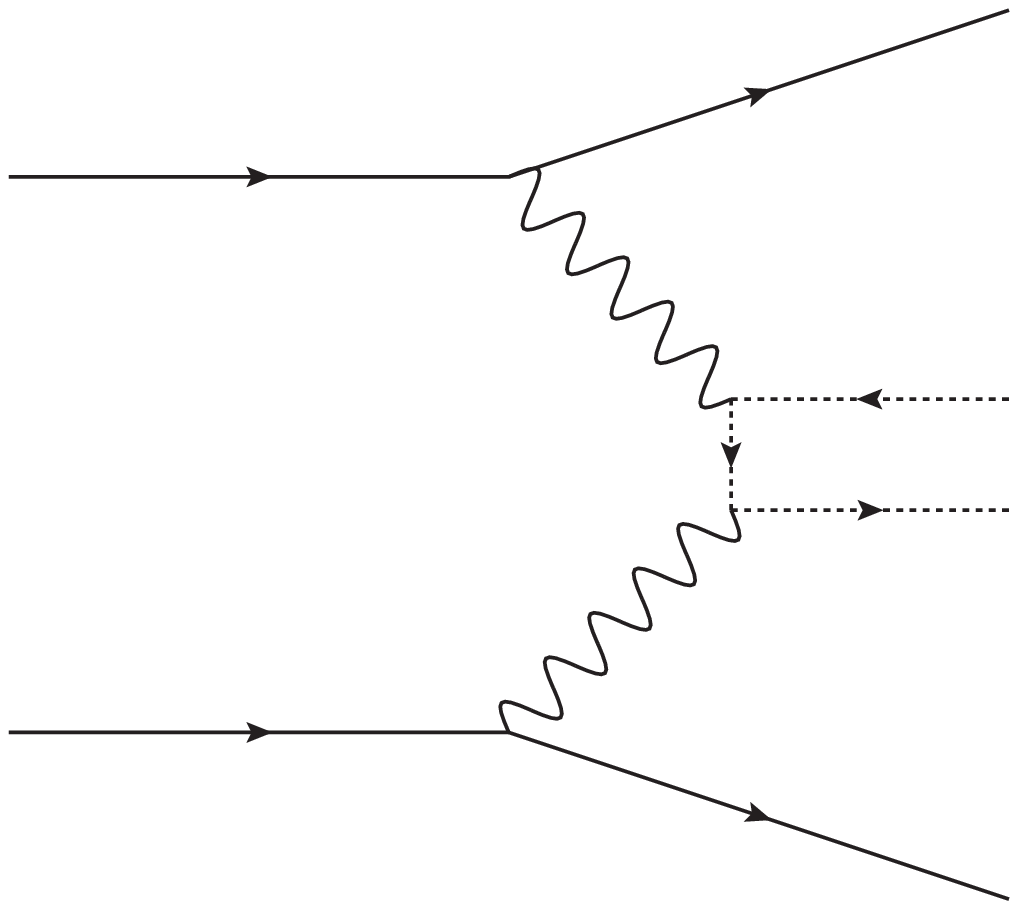} \includegraphics*[width=0.45\columnwidth]{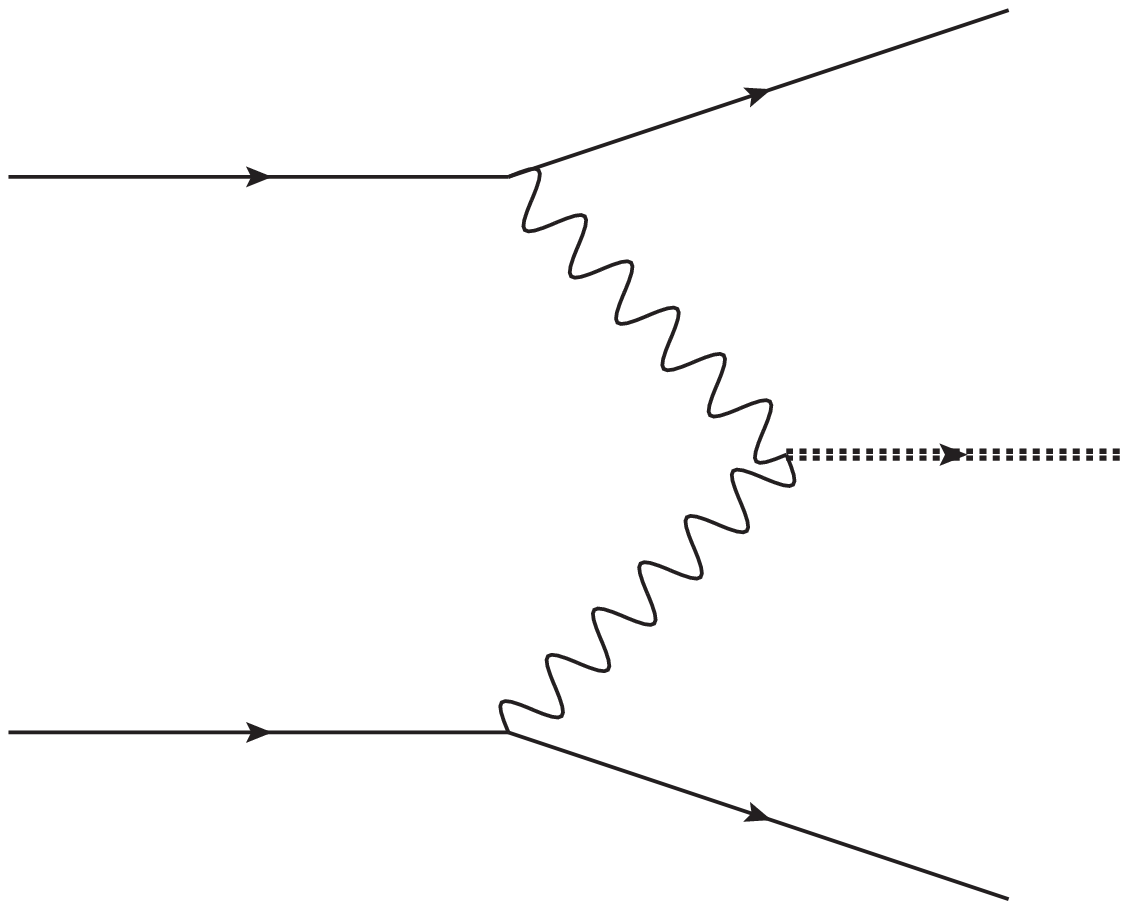} 
 \end{center}
 \caption{Production of monopole/antimonopole pair (left) and monopolium (right). The straight lines are the projectiles, the wavy lines, photons and the dotted lines, monopoles or monopolium.} \label{fig:ff}
 \end{figure}

The process can be described\cite{khoze2000CanHigbeseerapgapeveTevorLHC,khoze2001DoudifprohigresmismasexpTev,petrov2004ExcdoudifeveMenLHC} as the collision of two hadrons (or leptons) which interact themselves by the gauge boson exchange (see Fig. \ref{fig:ff}). In exclusive channels, the projectiles remain intact after the interaction and the gauge bosons combine, generating a massive resonance with the same quantum numbers of the vacuum, resulting in rapidity gaps in the detector.

This mechanism presents some advantages, as, for example, a very clean experimental signature, a improved mass resolution and  a suppressed background. For another side, there exist disadvantages: the theory is not free of divergences, the measure of cross sections is hard-working, requiring detectors installed away of the interaction point\cite{albrow2010CenExcParProHigEneHadCol} and the experimental signal is low.

Our focus is at photon processes in peripheral collisions\cite{baur2002Cohgamgamgamintverpercolrelioncol,bertulani2005Phyultnuccol,baltz2008PhyUltColLHC}. This process is described using the photon equivalent approximation. In this picture, a electric charged particle with high energy have the electromagnetic fields concentrated in his transverse region and can be substituted by a equivalent photon flux. This photons interact to produce the massive resonance. In peripheral collisions, the impact parameter is great than the sum of radius of the particles, avoid frontal collisions and thus high particle multiplicities produced in the interaction.

Using the most simple model \cite{nystrand2005Eleintnucprocol} for a estimation for total cross sections in all the cases: electron-electron, proton-proton and ion-ion collisions,
 \begin{equation} \label{eq:photequiv}
\sigma_\mathrm{tot} = \int_{M_{\gamma\gamma}^2/s_{NN}}^1 dx_1 f_A(x_1) \int_{M_{\gamma\gamma}^2/x_1s_{NN}}^1 dx_2 f_B(x_2) \sigma_{\gamma\gamma}(\hat{s})
 \end{equation}
where $M_{\gamma\gamma}$ is the mass of the central produced system, $s_{NN}$ is the center-of-mass energy of the projectiles, $x_i$ is the fraction of the energy of the photon $i$, $\hat{s} = x_1 x_2 s_{NN}$ and $f(x)$ is the photon energy spectrum produced by a charged particle. The Weizsacker/Williams\cite{williams1934Nathigparpenradstaionradfor} expression for photon spectrum (used for ion collisions) is
\begin{equation} \label{eq:ww}
 f_\mathrm{WW}(x) = \frac{\alpha_{elm}Z^2}{\pi}\frac{1}{x}\left[ 2YK_0(Y)K_1(Y) - Y^2\left(K_1^2(Y) - K_0^2(Y)\right) \right],
\end{equation}
with $Y=xM_Ab_{min}$ where $\alpha_{elm}$ is the fine structure caonstant, $Z$ is the atomic number of the projectile, $M_A$ is the mass of projectiles, $b_{min}$ is the minimum impact parameter and $K_i$ is the modified Bessel functions. For proton collisions, we use the Dress and Zeppenfeld photon spectrum\cite{drees1989ProSupParElaepCol} given by 
\begin{equation}
f_\mathrm{DZ}(x) =  \frac{\alpha_{elm}Z^2}{2\pi x}\left[1+(1-x)^2\right]\left[ \ln A - \frac{11}{6} + \frac{3}{A} - \frac{3}{2A^2} + \frac{1}{3A^3}\right]
\end{equation}
where 
\[ 
A = 1 + \frac{\unit{0.71}{\GeV^2}}{Q^2_\mathrm{min}},\qquad Q_\mathrm{min}^2 \simeq \frac{m_p^2x^2}{1-x}.
\]

 For the electron case, we use the expression of Frixione\cite{frixione1993ImpWeiappele-procol},
\begin{equation}
 f_{e}(x) = \frac{\alpha_{elm}}{2\pi} \bigg\{ 2m_{e}^{2}x \left[ \frac{1}{q^2_{max}} - \frac{1}{q^2_{min}}\right] + \frac{1+(1-x)^2}{x}\log\left(\frac{q^2_{min}}{q^2_{max}}\right) \bigg\}
\end{equation}
where 
\[ 
 q^2_{max} = -\frac{m_{e}^{2}x^2}{1-x},\quad q^2_{min} = -\frac{m_{e}^{2}x^2}{1-x} - E^2(1-x)\theta_c^2
\]
with $m_e$ is the electron mass and $E$ is the energy beam and $\theta_{c} = \unit{30}{mrad}$.

Same question are address in this process: the rise of the photon flux with atomic number of projectiles ($Z^4$); the low luminosity in ion-ion collisions; the Coulomb dissociation and excited states of projectiles \cite{hencken1995PholumrelheaioncolLHCene,baltz1998SupheaiongamgamproHigbyCoudis} (not consider here) and the nuclear charge form factor which modifies the photon flux and the overlap of hadron tails in the collision.

\section{Results \label{sc:res}}

\begin{figure}[ht]
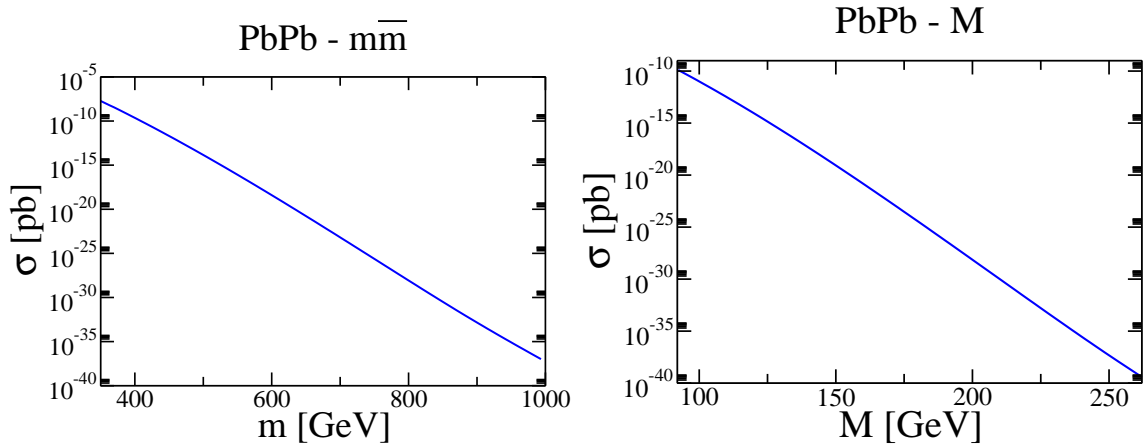
 
 \begin{center}
 \includegraphics*[width=0.45\columnwidth]{mm-chumbo} \includegraphics*[width=0.45\columnwidth]{M-chumbo} 
 \end{center}
 \caption{Total cross section for the production of monopole-antimonopole (left) and monopolium (right) in lead-lead collisions at LHC energies as function of the mass of monopole/monopolium.} \label{fig:pb}
 \end{figure}

We calculate the cross section as a function of the mass of central system at fixed center-of-mass energies corresponding to different colliders. First, we present in the Fig. (\ref{fig:pb}) the results of lead-lead collisions at LHC energy ($\sqrt{s} = \unit{5.5}{\GeV}$) for the production of monopole/antimonopole pairs and monopolium. As expected, the cross section decreases with the raise of the mass and the values are very low, although the enhancement in atomic number ($Z^4$) in the photon luminosity of the projectiles. As the massive central system requires a great amount of energy, there are very few energetic photons produced by a ionic projectile, diminish the cross section.

\begin{figure}[ht]
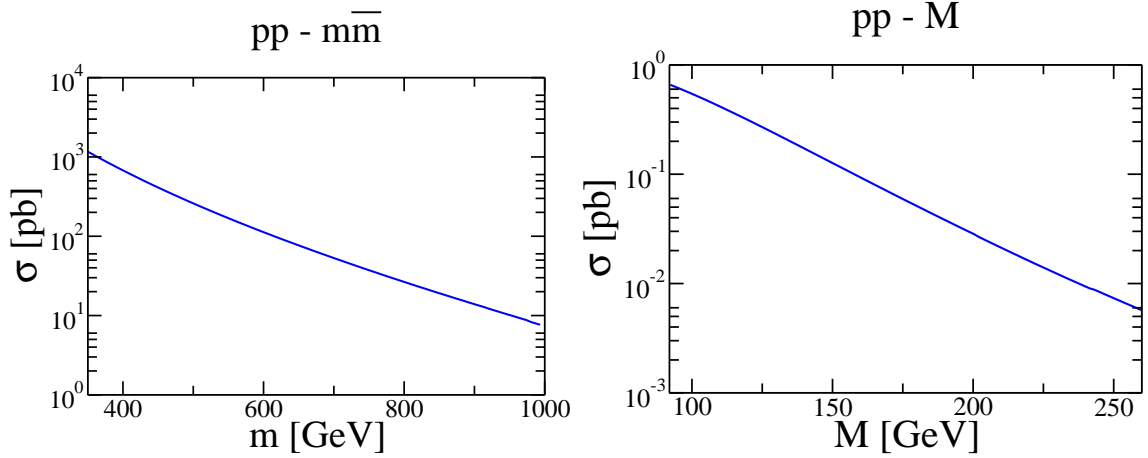
 
 \begin{center}
 \includegraphics*[width=0.45\columnwidth]{mm-proton} \includegraphics*[width=0.45\columnwidth]{M-proton} 
 \end{center}
 \caption{Total cross section for the production of monopole-antimonopole (left) and monopolium (right) in proton-proton collisions at LHC energies as function of the mass of monopole/monopolium.} \label{fig:pp}
 \end{figure}

Next, we calculate the same quantity for proton-proton collisions in the LHC energy ($\sqrt{s} = \unit{14.0}{\GeV}$) and displayed the results in the Fig. (\ref{fig:pp}). The above general features of the result are the same as in lead-lead collision case, except that the cross section in this case are greater than the previous one. Comparing different produced particles scenarios, the monopolium have a cross section three orders of magnitude smaller. For comparison with the previous results\cite{kurochkin2006promagmonviagamgamfushigeneppcol,dougall2009Dirmagmonprophofusprocol,epele2012LoomagmonLHCdipeve}, the present one for both cases are agree with the previous results found in the literature.

At last, we present the results of cross section for the production of monopoles and monopolium at electron-positron collisions at future CLIC energies\cite{Aicheler:2012bya}. Again, in comparison with the results from hadron projectiles, we obtain a same behavior as a function of the central produced system but with a significant larger cross section. As in previous cases, due the large mass of monopolium, his cross is smaller than the monopole/antimonopole case.

\begin{figure}[ht]
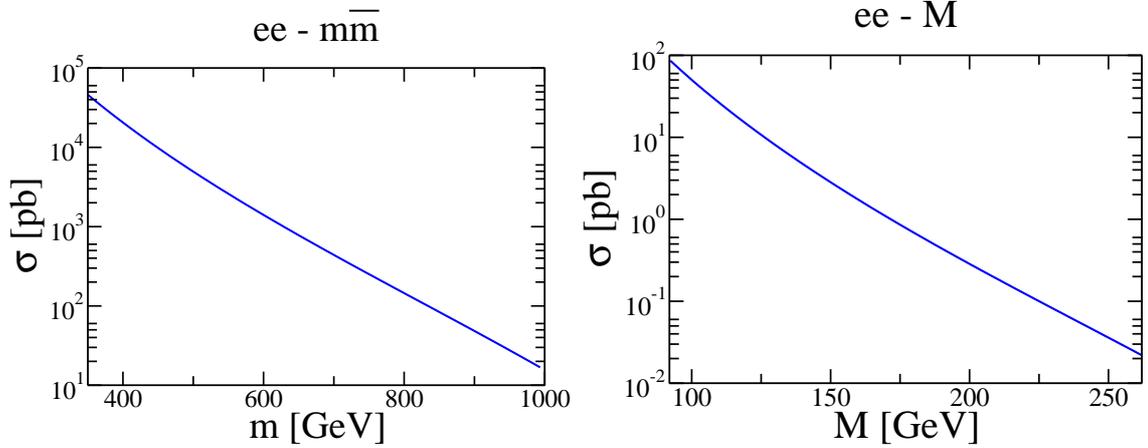
 
 \begin{center}
 \includegraphics*[width=0.45\columnwidth]{mm-eletron} \includegraphics*[width=0.45\columnwidth]{M-eletron} 
 \end{center}
 \caption{Total cross section for the production of monopole-antimonopole (left) and monopolium (right) in electron-electron collisions at planned CLIC energies as function of the mass of monopole/monopolium.} \label{fig:ee}
 \end{figure} 

\section{Summary and conclusions \label{sc:cls}}

In this work, we consider the central exclusive production of (Dirac) magnetic monopoles and the bound state of a monopole and antimonopole, the monopolium, for hadronic and eletronic collissions at LHC and (planned) CLIC energies. We present a prediction for the cross sections for lead-lead collisions and electron-electron collisions and, for the proton case, a comparison with the previous results.

In the treatment of the production of magnetic monopoles exist several drawbacks. One of them is the strong coupling with photons, which not justificate a perturbative calculation. Another problem is absence of direct experimental observation of magnetic monopoles. In the literature, only estimatives of mass and charge can be found. Other disadvantage is the large mass of the central state, already delimitaded by the experimental results avaliable. In present formalism, the production of this central states are disfavored due the very small number of equivalent photons with required energy to produce this particles. In the lead case, we have a large radii and, as we interessed in exclusive processes, the number of photons rise quickly, nullyfing the gain in atomic number $Z$ in comparisson with proton collisions and also have a low luminosity in the collider.

However, besides the above issues, the results are promissing in the case of electronic and proton collisions. In particular, the productiion in a electron-electron collider will be measurable in a significative rate of events, basead in the planned luminosity of the CLIC collider and the above results for the cross section in this case, for a large gap of monopole mass values and even in the case of exclusive production, which have a small cross section. Comparing the proton and electron processes, the first one could include inclusive and inelastic processes, that rises the total cross section, while the electron collisions only have the process consider in this work, which gives a very clean experimental signal.

\begin{acknowledgments}
The authors thanks the Grupo de Altas e M\'edias Energias for the support in all stages of this work. J.T.R. thanks CAPES for the financial support during the development of this work. 
\end{acknowledgments}

\bibliography{referencias}

\end{document}